\documentclass[letterpaper,10pt,twocolumn]{ASPE_ExtendedAbstract}
\usepackage{lipsum}
\usepackage{siunitx}
\usepackage{subfigure}
\usepackage{url}

\title{An Experiment to Test the Mechanical Losses of Different Bonding Techniques in Fused Silica}

\author{Jonathan J. Carter$^{1,2}$, Pascal Birckigt$^{3,4}$, Oliver Gerberding$^{5}$,  Qingfeng Li$^{4}$,\\Rick Strüning$^{2}$, Tobias Ullsperger$^{4}$, and Sina M. Koehlenbeck$^{1,2}$}

\affiliations{1 Max Planck Institute for Gravitational Physics (Albert Einstein Institute), \\ Callinstraße 38, 30167 Hannover, Germany\\
2 Institut für Gravitationsphysik der Leibniz Universität Hannover,\\ Callinstraße 38, 30167 Hannover, Germany\\
3 Fraunhofer Institute for Applied Optics and Precision Engineering,\\ Albert-Einstein-Str. 7, 07745 Jena, Germany\\
4 Institute of Applied Physics, Abbe Center of Photonics,\\ Friedrich-Schiller-Universität Jena, Albert-Einstein-Str. 15, 07745 Jena, Germany
\\5 Institut f\"{u}r Experimentalphysik,\\
Luruper Chaussee 149, 22761 Hamburg, Germany}

\usepackage{graphicx}

\usepackage{amsmath}
\usepackage{amssymb}

\intextsep=\lineskip

\begin{document}
\maketitle 

\section*{Abstract}  
High-purity glasses are used for their low optical and mechanical loss, which makes them an excellent material for oscillators in optical systems, such as inertial sensors. Complex geometries often require the assembly of multiple pieces of glass and their permanent bonding. One common method is hydroxide catalysis bonding, which leaves an enclosed medium layer. This layer has different mechanical properties to the bulk glass around it. The higher mechanical loss of this layer makes it more susceptible to displacement noise originating from the conversion of energy from oscillation to heat and vice versa.  Therefore, other methods are needed to bond together glass assemblies. To investigate this, we have set up an experiment to measure the mechanical losses of several different types of bond commonly used in fused silica manufacturing, namely; plasma activated direct bonding, hydroxide catalysis bonding, laser welding, and adhesive bonding. In this paper we present the experimental design and show initial results of the first test sample.

\section{Introduction}
Intrinsic thermal noise is a limit to the performance of many high precision optomechanical devices \cite{Zhou2021}. 
Via the fluctuation dissipation theorem, the thermal noise is directly linked to the mechanical dissipation of the device. 
To prevent additional loss channels, many optomechanical measurement devices are constructed monolithically \cite{Cumming2012,Bellouard2005}.
One-piece constructions can be impractical, and complex components must then be assembled from multiple parts due to their geometry and size \cite{Carter2020}. 
There are now a wide variety of glass bonding techniques available to construct quasi-monolithic assemblies, such as optical contacting and hydroxide catalysis bonding. 
The associated losses, in particular of hydroxide catalysis bonding \cite{Smith2003,Cunningham2010}, have been studied in the course of the construction of the advanced Laser Interferometer Gravitational-Wave Observatories (aLIGO). Others, such as plasma activated direct bonding \cite{Shari1995}, have no measurements of their intrinsic losses in the literature to the authors' knowledge. 

This paper details work ongoing to measure the mechanical losses of several fused silica bonding techniques. The production process, manufacturing of the test samples, and the experiment to measure he mechanical Quality factor, $Q$, are detailed, as well as results from the first tested sample.

\section{Micro-Oscillator Design}
\begin{figure}[t]
    \centering
    \includegraphics[scale=.6]{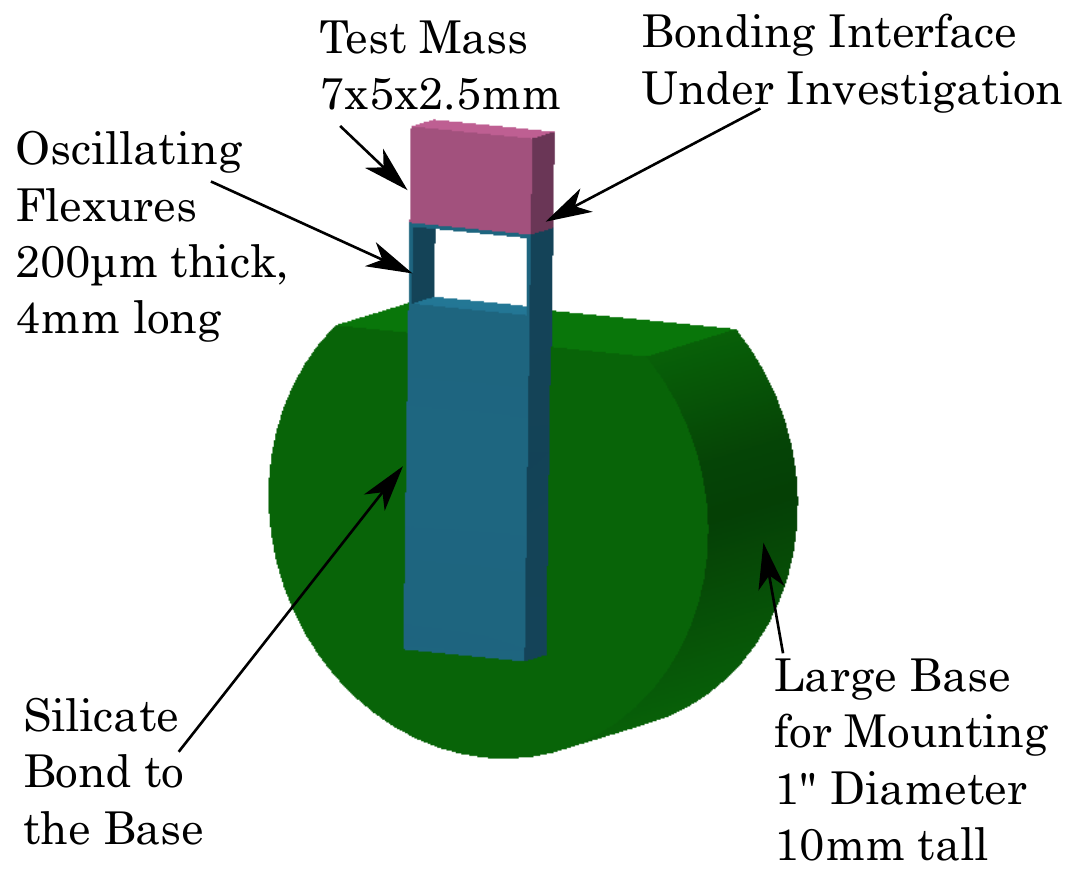}
    \caption{The test sample design. There are three separate glass parts depicted in green, blue and violet. The large round base, in green, is attached to the flexure piece, in blue, via silicate bonding. This bonding interface is necessary, but not under investigation, as it is far from any oscillating parts. The flexure piece is attached to the test mass, in violet, at a distance of \SI{200}{\micro\meter} from the flexures. This connection is created with the bonding technique under investigation. In the case of monolithic control samples, there is no bonding interface here, but bulk glass.  The flexure piece uses a parallelogram geometry to oscillate in plane.}
    \label{fig:bondingtestPieceScheme}
\end{figure}
A micro-oscillator with bonding interfaces in key locations was designed for this experiment. The $Q$ factor is related to the mechanical loss angle, $\phi$ by
\begin{equation}
    Q=\frac{1}{\phi}.
\end{equation}
By measuring the $Q$ factor, and understanding all loss terms, it is possible to infer what contribution is due to the bonding interface itself.  

Our device design is shown in FIGURE \ref{fig:bondingtestPieceScheme}. It takes inspiration from work by Guzmán et al. \cite{Guzman2014}, who showed that a device using parallelogram flexure structures can make for a high $Q$ factor oscillation along one spatial direction. The bonding interface between two fused silica parts are varied between the bonding techniques listed in a separate section below.

The bonding surfaces is located in the top part of the assembly, between the test mass and the flexure  piece. As this part is only held up by glass flexures of \SI{200}{\micro\meter} thickness, it is able to move with respect to the base. The flexures are bent by the oscillation which induce stress along their length and in the corners of the base and test mass. The bonding surface is as close as possible to this deformation induced stress, thus contributing a loss channel and lowering the $Q$ factor of the oscillation. 

 Finite Element Analysis (FEA) with COMSOL Multiphysics was used to analyse the oscillations as well as the induced stress in the design.
It was found that the first eigen mode was a horizontal motion 
 at approximately \SI{2.1}{\kilo\hertz}. This is the mode desired and excited during measurements of the $Q$ factor. A second, vertical mode occurs at \SI{5.1}{\kilo\hertz}. This is undesired, but as it is more than a factor of two above in frequency and orthogonal, cross contamination is expected to be low. The next modes occur in the tens of \si{\kilo\hertz} and are all rotational. These are assumed to be too high in frequency to interfere with the measurement.

\begin{figure}
    \centering
    \includegraphics[scale=.55]{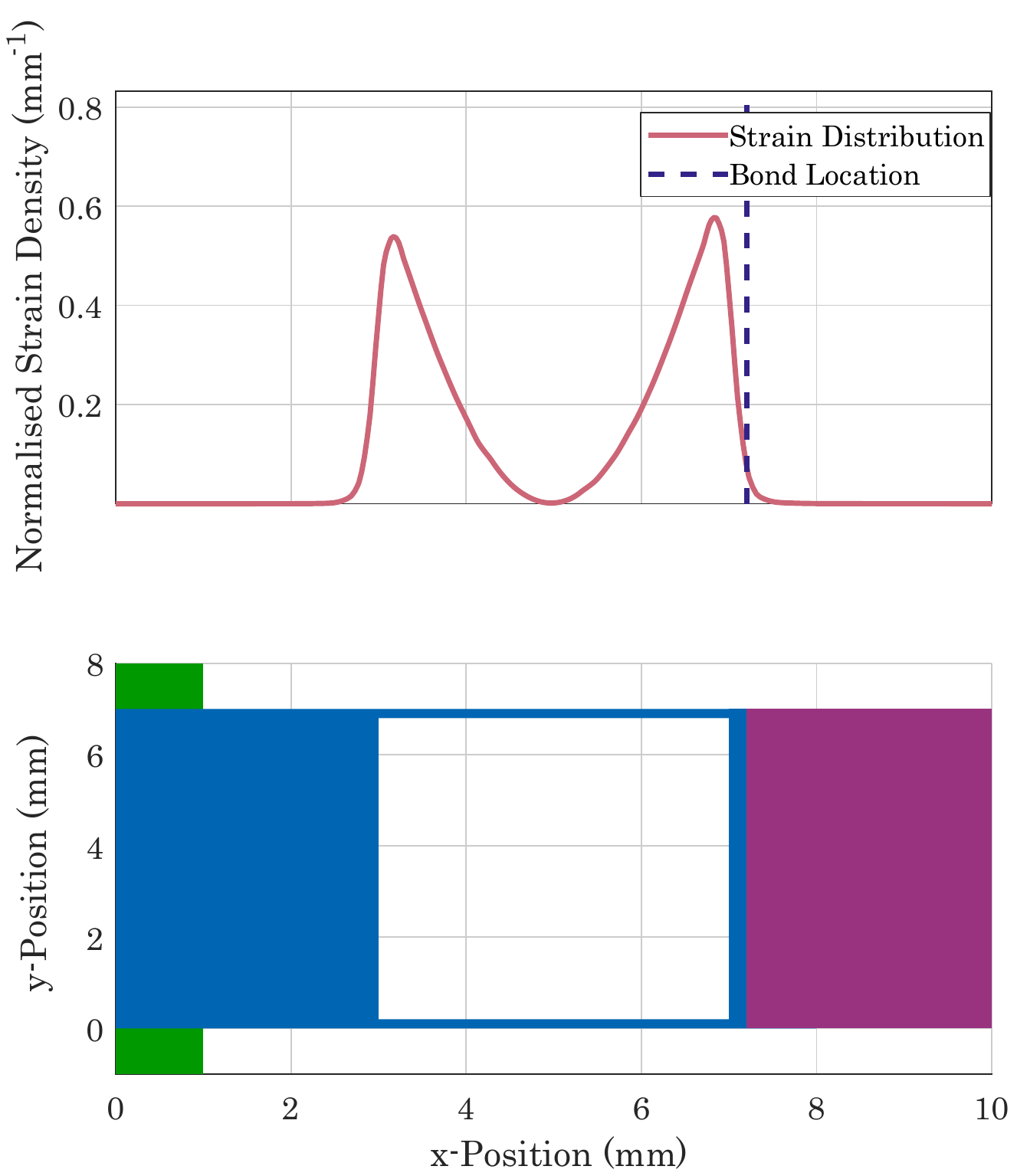}
    
    \caption{ Normalised ratio of strain energy distribution across the geometry estimated by FEA.  The strain distribution peaks  in the corners of the oscillating part and decaying rapidly as it disperses into the bulk. The location of the bonding surface is shown with the dashed line and is as close as possible to the flexures to maximise strain across the bonding interface. }
    \label{fig:straindis}
\end{figure}

FEA is also used to measure the strain energy distribution across the flexure elements. This is necessary for two aspects, the first is estimating the contribution of the material effect to the $Q$ factor of the oscillator, the second is for calculating the energy in the bonding interface and so derive its actual loss coefficient. The distribution of strain energy was simulated by applying a fixed extension on the test mass and simulating how the strain energy distribution changed in segments along the length of the flexures. This was then normalised with the total strain energy in the system to produce FIGURE \ref{fig:straindis}.

The leading limiting source of loss in the design is from the material itself which we will describe with its respective $Q$ factor  $Q_{\mathrm{{Mat}}}$. This term can be subdivided further into bulk and surface loss. Bulk loss is the loss in the volume of the oscillator intrinsic to the material it is made from. As such, the experiment uses a low loss Corning 7980-0F fused Silica with an intrinsic $Q$ factor of $1.1\times10^{7}$  as measured by \cite{Numata2001}. 

Surface loss can be understood in a similar manner to bulk loss. At the surface of the oscillator damage occurs that breaks the regular structure and allows contaminates to penetrate in the material acting as an additional loss source. This leads to an understanding of a low loss bulk with a surface layer of depth, $d_{\rm{s}}$, with a higher loss. The associate $Q$ factor can be described mathematically as 
\begin{equation}
	\frac{1}{Q_{\mathrm{{Mat}}}} = \phi_{\text {Bulk }}+\phi_{\text {Surf }} d_{\rm{s}} \frac{\int_{S} \epsilon^{2}(\vec{r}) \mathrm{d} S}{\int_{V} \epsilon^{2}(\vec{r}) \mathrm{d} V},
    \label{eqn:bulkloss}
\end{equation}
where $\phi$ is the loss angle in the bulk and surface layer respectively, and $\epsilon$ is the strain energy density.
Gretarsson et al. \cite{Gretarsson1999} made measurements of some of these factors in the late 90s. This estimated the approximate contribution $\phi_{\rm{Surf}}$ as $10^{-5}$ and a $d_{\rm{s}}$ of \SI{1}{\micro\meter}. Although these values depend heavily on the treatment of samples, we assume it to be comparable.
Substituting in values from FIGURE \ref{fig:straindis} gives a $Q_{\rm{Mat}}$ of \SI{9e6}{}. 

Contributions to the $Q$ factor from thermoelastic damping, $Q_{\rm{TED}}$, can be estimated using the Zener approximation. By following the method of Lifshitz \cite{Lifshitz1999}, an estimated contribution of $Q_{\rm{TED}}$ as \SI{2e7}{} is reached for this design. The contribution of Akheizer loss, see \cite{Nawrodt2013} for details, is also very similar with $Q_{\rm{ph-ph}}$ of \SI{2e7}{}. Air pressure damping is minimised by performing the tests under vacuum, leaving only frictional losses as the unknown loss term. As this does not belong to the sample itself, we can estimate the theoretical limit of the $Q$ factor of this oscillator design as \SI{4.5e6}{}.

\begin{figure}[t]
\centering
\includegraphics[scale=.3]{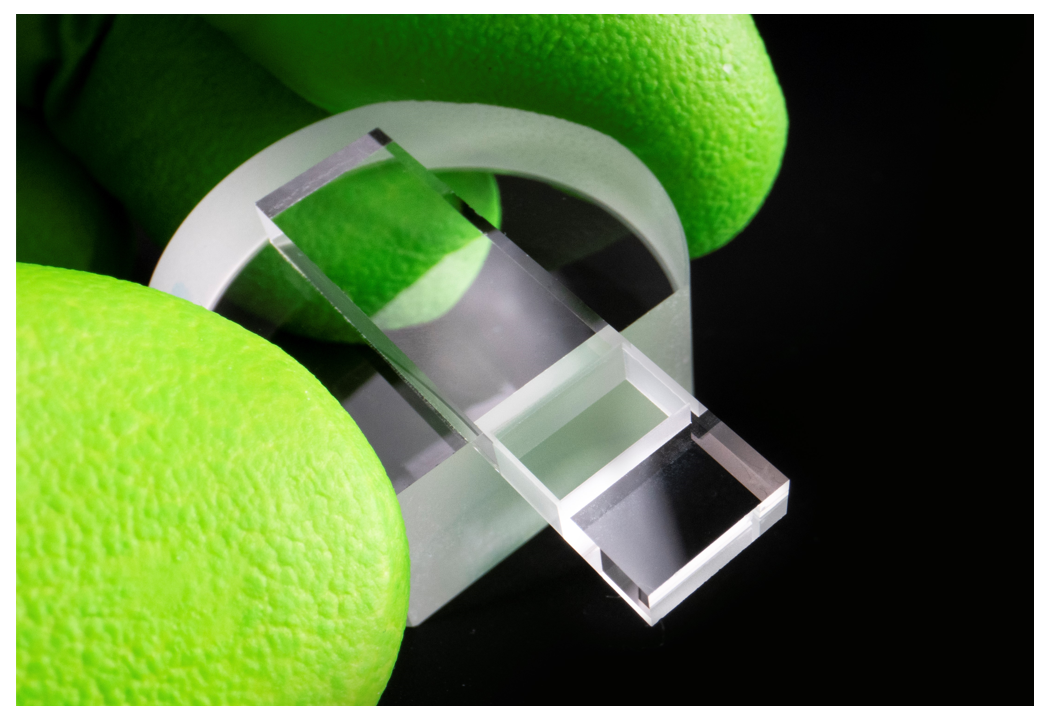}
\caption{Photo of an exemplary test sample.}
    \label{fig:sample-photo}
\end{figure}

Several samples of each bonding technique have now been produced and are ready for testing. An exemplary sample is shown in FIGURE \ref{fig:sample-photo}. In addition, several devices with no bonding interface, i.\,e., monolithically, were produced, which act as a reference for the bonded samples.

\begin{figure}
    \centering
    \includegraphics[width=0.45\textwidth]{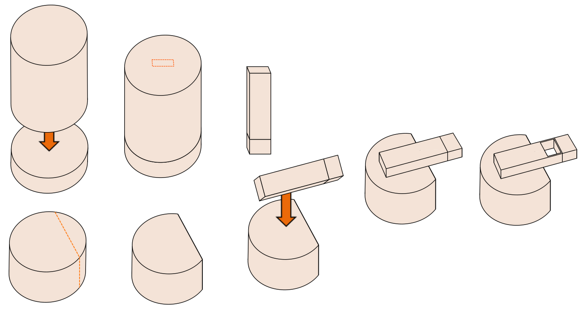}
    \caption{Schematic manufacturing steps for producing the test samples. Firstly, two round plates are joined using a certain bonding technique. Then, it is cut into  cuboids via diamond sawing, which are then carefully polished from all sides. Meanwhile, a segment of another  round plate is cut off to form the base. The cuboid is joined with the base.  Finally, laser ablation is applied using an ultrashort pulsed laser.
    }
    \label{fig:sample-manufacturing}
\end{figure}

All samples were produced from double-side polished cylindrical pieces of \textsc{Corning} 7980-0F fused silica \SI{1}{}" in diameter round plates. The sample manufacturing steps are shown in FIGURE \ref{fig:sample-manufacturing}. 
As a final manufacturing step, the thin flexures were exposed by precise micro-machining using an ultrashort pulsed laser system, TruMicro5070. Starting from the backside of the cuboid, the focused laser pulses were scanned across the whole rectangular area except of the thin flexures at the edges. During the layer-wise ablation process the focal position was continuously moved up to realise taper-free cuts, leaving flexible elements of only \SI{200}{\micro\meter} width.


\section{Bonding Techniques}
\label{sec:bonding}
This section describes the various joining techniques for fused silica pieces under investigation ranging from simple gluing to glass welding. 

\subsubsection{Hydroxide Catalysis Bonding}
Hydroxide-catalysis bonding, also called silicate bonding, is a form of bonding between glass surfaces invented by Gwo et al. at Stanford \cite{Gwo1998}. It has been used extensively in the gravitational wave community because of its low mechanical loss and high strength and so is well studied \cite{Smith2003,Cunningham2010,Rowan1998}. The bonding process uses two polished, flat surfaces of silica based glass. One surface is treated with a hydrous solution, here a sodium silicate solution, then, the other surface is put into contact. During curing, the solution forms long siloxane chains. After the water has evaporated, these interact and tangle together forming a strong, highly stiff interlayer  \cite{Veggel2008}.
\subsubsection{Adhesive Bonding}
Adhesive bonding is a broad term used to describe any bonding method with an organic glue as medium between the two surfaces. In this specific case, Epotec 301-2 was used. While this is the least promising bonding technique in terms of thermal noise performance, it is the simplest bonding technique. Knowing what level of thermal noise this bonding can achieve is therefore of broad interest to the community.
\subsubsection{Plasma Activated Bonding}
Plasma activated bonding, which is a low-temperature direct process, is performed by optically contacting two clean, highly polished, highly flat surfaces. Surfaces are prepared via chemical cleaning and a plasma-activation step. After contacting, a force of several \si{\kilo\pascal} is applied.  Then, a process of thermal annealing converts the resulting hydrogen bonds  into strong covalent bonds, making a stable interface \cite{ Kalkowski2011,Tan2006}. As direct bonding yields a high-strength, interlayer-free interface, it is perhaps the most promising bonding technique in terms of thermal noise performance. 
\subsubsection{Ultrashort Pulse Laser (USP) Glass Welding}
In USP glass welding, the lasers have a wavelength to which the glass is transparent. Significant energy input can only be established via nonlinear absorption, e.g. multiphoton ionisation, when high intensity, of order hundreds of \si{\tera\watt\per\square\centi\meter}, is reached by tight focusing.  When two glass surfaces are brought together and the laser beam is focused at the interface, the glass near the focal volume is melted across the interface so that a new joint is formed. Here, the melting points are placed on a grid with a distance of \SI{50}{\micro\meter}. More details are given in \cite{Richter2012}. Previous studies have investigated the mechanical loss of this bonding technique and found it to be less than the surface loss of their tested samples and thus was not detectable \cite{Harry2006}. 

\section{Experiment}
In order to measure the $Q$ factor of the samples, two things are needed, a way to excite motion and a way to measure the samples response. A piezo element attached to the mount of the oscillator can do the former. As the device is made of optical grade fused silica, an interferometer can do the later.

\begin{figure}
    \centering
        {\includegraphics[scale=.6]{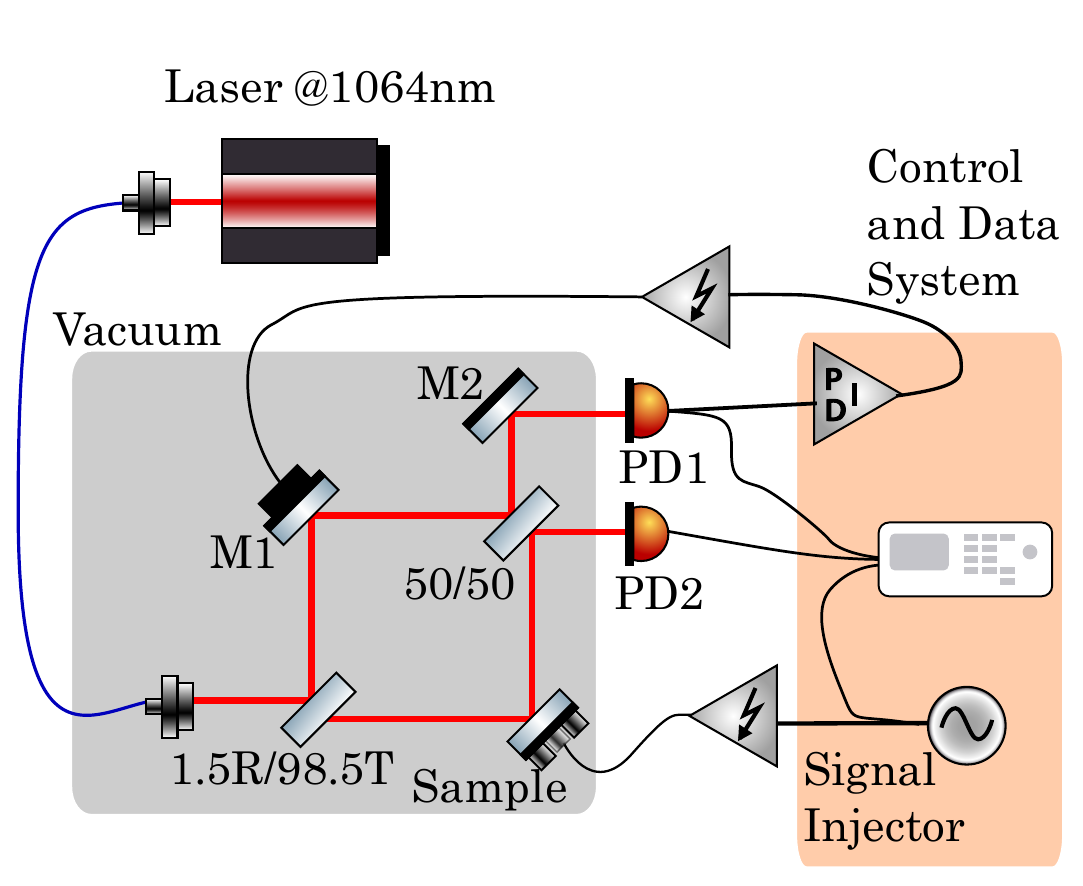}}
        \caption{Left: The experimental schematics of the interferometer used to test the samples. The experiment uses laser light from at \SI{1064}{\nano\meter} laser. This is sent onto a breadboard in vacuum via optical fibre feedthroughs. It is coupled out and send into Mach Zehnder interferometer. The power is not distributed evenly, but rather via a \SI{1.5/98.5}{} beamsplitter to compensate for the low reflectivity of fused silica. M1 is driven by a controller to keep the interferometer at the operating point, which is the mid-fringe. The sample is in a custom mount attached to a shear-piezo which is driven allowing for a measure of its transfer function. 
    }
    \label{fig:bondingtestscheme}
\end{figure}

The interferometer is a homodyne interferometer in a Mach Zehnder configuration, as shown in FIGURE \ref{fig:bondingtestscheme}. A piezo driven mirror in one arm keeps the interferometer on the mid fringe operating point, ensuring a linear response of the interferometer readout. A shear-piezo mounted under the sample causes excitation for measurement of the mechanical response of the oscillator.

After setting up the experiment in a vacuum environment, the first step was to measure the background noise of the interferometric readout. For this purpose, we used a device that was shaped like the test pieces but had no bending geometry, allowing it to remain a rigid piece. It was found that there were no significant resonance peaks in the region of interest, and a noise floor of \SI{1e-11}{\meter\hertz\textsuperscript{-1/2}} was achieved.

\section{Results \& Discussion}
The first sample under test was one of the monolithic reference samples with bending flexures. Since no loss channel from a bond is present this will give the intrinsic loss of the sample material and geometry and the experimental set-up. 

\begin{figure}
    \centering
    \includegraphics[scale=.55]{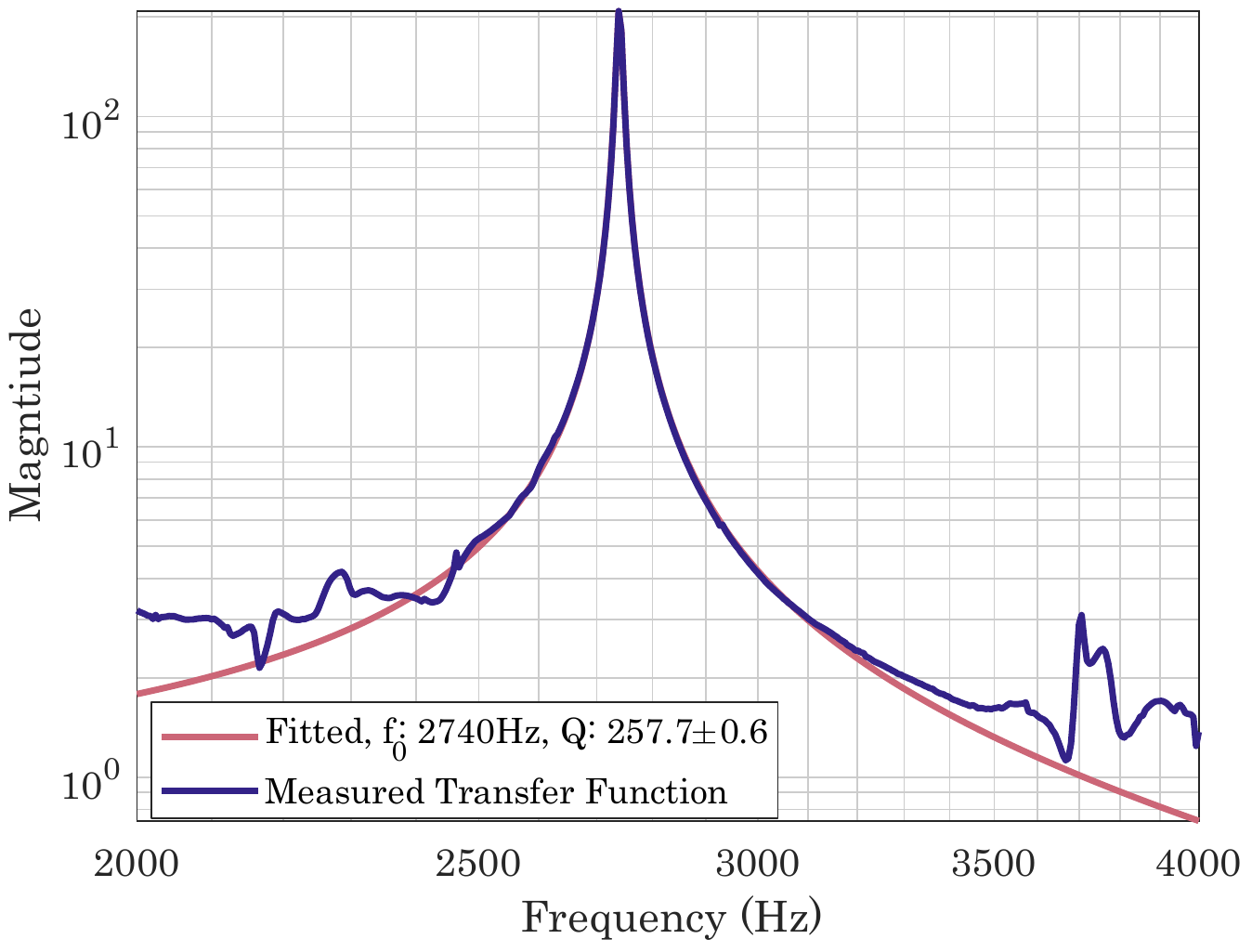}
    \caption{The fit and measured absolute transfer function of monolithic sample 13. The peak location has a deviation of 30$\%$ from  the value predicted by FEA simulation. 
    As the resonance frequency is strongly dependent on flexure thickness the difference is mainly attributed to the small variations in thickness seen along the flexures.}
    \label{fig:monopeakfit}
\end{figure}

An excitation signal was applied to the shear-piezo, driving the sample across Fourier frequencies from \SIrange[]{2}{4}{\kilo\hertz}. The displacement of the test mass was recorded with the interferometer and the transfer function from the driving displacement to the test mass displacement was calculated. The amplitude of the transfer function was then fit with the model of a simple damped harmonic oscillator, given by
\begin{equation}
    \Tilde{X}=\left|\frac{\omega_{\rm{0}}^2}{\left(-\omega^2+{\omega_0}^2+\frac{i\omega_0\omega}{Q}\right)}\right|,
    \label{eqn:tfseismo}
\end{equation}
where $\omega_{\rm{0}}$ is the natural angular frequency, and $\Tilde{X}$ is the magnitude of the frequency response to stimulation. The fit was done using MATLABs Nlinfit function. The result of this fit is shown in FIGURE \ref{fig:monopeakfit}.

The $Q$ factor is \SI{257.7}{}, three orders of magnitude below where it is expected to be. The cause of the discrepancy is unknown, though there are two ideas that are being investigated. The first is that contaminants had entered the flexures. There is also some concerns that the largely unknown term of frictional loss are limiting the measurements.  Alternatives to the currently used clamp are being investigated. It is hoped that improving the rigidity of the clamp will lead to less differential motion and so less friction. 

\section{Conclusions}
This paper reports ongoing efforts to measure the mechanical loss intrinsic to several bonding techniques commonly used in fused silica manufacturing. 

A mechanical oscillator geometry has been designed for the purpose of measuring the loss in bonding interfaces. The oscillator design has a promising theoretical $Q$ factor and artificially induced loss channel to measure the bonding interfaces. Several test pieces have been produced. By using interferometry, a means of measuring the $Q$ factor of the sample has been achieved. This will in turn allow for an estimation of the mechanical loss due to the bond. However, the $Q$ factor of the substrate is not what it is expected to be. There are several reasons this could be the case and ongoing investigations seek to answer why.

Overall the experiment still has promise to produce measurements of the mechanical loss of several in glass bonding techniques. By knowing this, larger fused silica glass assemblies can be made, without being limited by thermal noise. It is hoped that this will prove valuable information both to the specific field of gravitational wave astronomy and the broader glass optomechanical sensor design field as a whole.

\section{Acknowledgements}
We would like to thank Prof. Dr. Gerhard Heinzel, Prof. Dr. Benno Willke, Prof. Dr. Uwe Zeitner, Dr.-Ing. Stefan Risse, and Dr. Harald L{\"u}ck for their  continuing support of the project and are gratefully for their advice and ideas.

This project was supported by the Max Planck and Fraunhofer Institute Cooperation "High-QG".


\end{document}